\documentclass[useAMS,usenatbib]{mn2e}

\usepackage{graphicx}
\usepackage{natbib}
\usepackage[dvipsnames]{color}
\usepackage{url}

\def\blue{}

\newcommand{\km}{\,\mbox{km}\,\mbox{s}^{-1}}
\newcommand{\cm}{\,\mbox{cm}^{-2}}
\def\Ha{H$\alpha$}

\title[Observations of a strange elliptical bubble  in NGC 6946]{Imaging and spectroscopic     observations of a strange elliptical bubble  in the northern arm of  the spiral galaxy NGC 6946}

\author[Efremov \& Moiseev]
{Yuri N. Efremov$^1$\thanks{efremov@sai.msu.ru} and Alexei V. Moiseev$^{1,2}$\\
  $^1$Sternberg Astronomical Institute of the Lomonosov Moscow State University, Moscow 119992, Russia\\
  $^2$Special Astrophysical Observatory, Russian Academy of Sciences,   Nizhnii Arkhyz 369167,  Russia\\
}

\begin{document}

\date{Accepted ....  Received ....}

\pagerange{\pageref{firstpage}--\pageref{lastpage}} \pubyear{2011}

\maketitle

\label{firstpage}

\begin{abstract}
NGC 6946, known as the Fireworks galaxy because of its high supernova rate and high star formation, is embedded in a very extended HI halo. Its northern spiral arm is well detached from the galactic main body. We found that this arm contains a large ($\sim300$ pc in size) Red Ellipse, \blue{named according to a strong contamination of the \Ha\, emission line on its optical images}. The ellipse is accompanied by a  short parallel  arc  and a few others still smaller and less regular;  a bright star cluster is seen inside these features. The complicated combination of arcs seems to be unique, it is only a bit similar to some SNRs. However,  the long-slit spectral data obtained  with the Russian  6-m telescope  did not confirm the origin of the nebula as a result of a single SN outburst. The emission-line spectrum corresponds to the photoionization by young hot stars with a small contribution of shock ionization.   The most likely explanation of the Red Ellipse is a superbbuble created by a collective feedback of massive stars in the star cluster located in the NE side of the Red Ellipse. However, the very regular elliptical shape of the nebulae seems  strange.
\end{abstract}


\maketitle

\section{INTRODUCTION}

NGC 6946, a nearby starburst spiral galaxy, also known as the ``Fireworks galaxy''  because of its prolific supernova production, is embedded in an enormous HI halo (Pisano 2014). The galaxy has a high star formation rate and a number of unusual star-forming regions, such as enigmatic Hodge complex which contains a young super star cluster \citep*{Larsen2002, Efremov2004,Efremov2007}.  NGC 6946  is famous  also for hosting the record number of observed SNe.

{We found that  apart from the above mentioned peculiarities, this galaxy  hosts one  more strange object,   which  in common images  is seen as just one of  its many HII regions.  However, in deep high-resolution images this object  is really wonderful.   One of the best existing optical picture  of NGC 6946 is  the data, obtained with  the Suprime-Cam on the Subaru 8.2-m telescope.}  Using this beautiful image from the Subaru telescope press-release, one of us (Yu.E.) has found  an unusual red ellipse inside the isolated northern spiral arm; this color might be caused by \Ha\, emission,  what we do establish in this paper.

\begin{figure*}
\includegraphics[width=\textwidth]{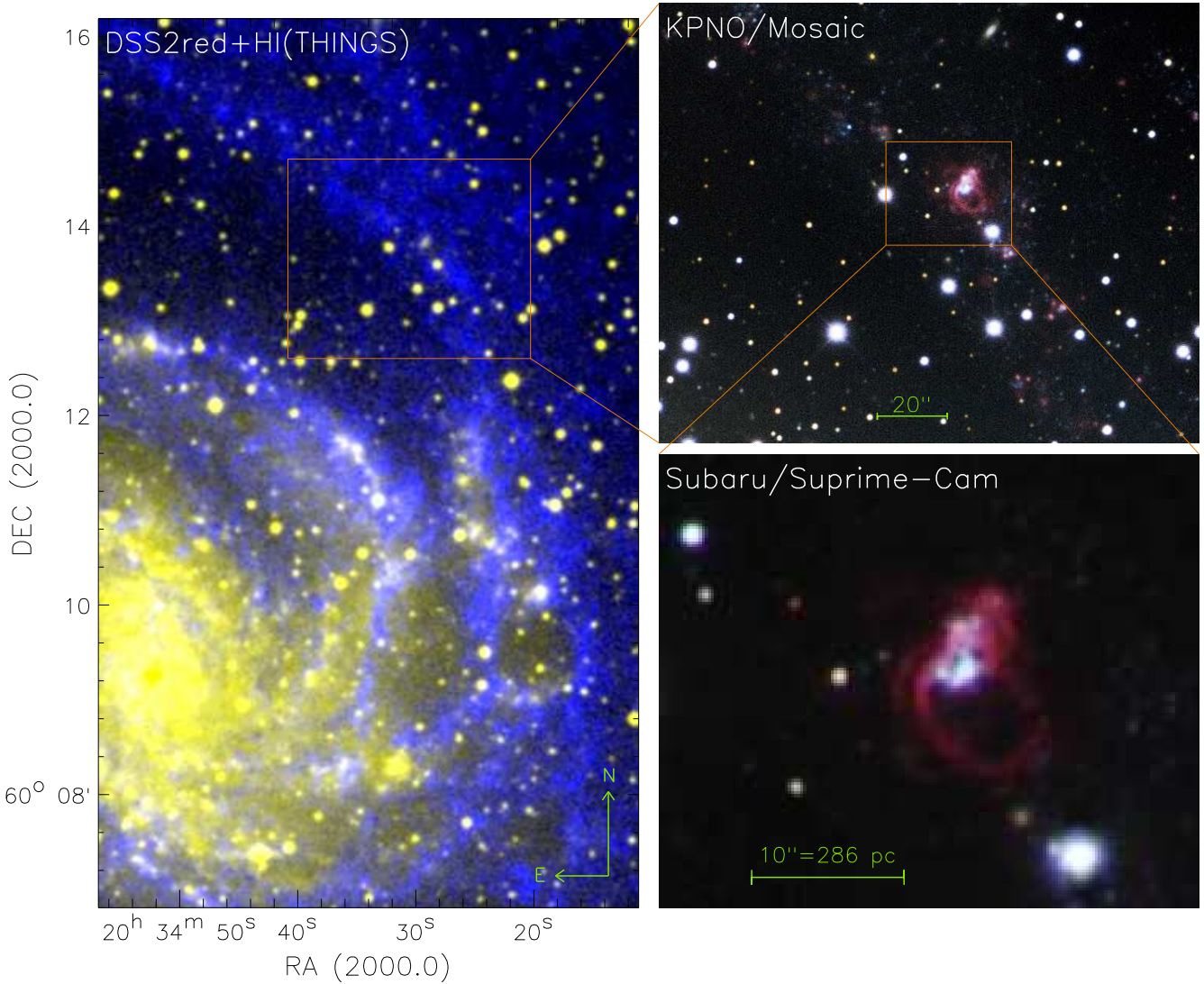}
\caption{Left: Composite optical (DSS2---white and yellow colors) plus  HI surface density map \citep[blue color, ][]{Walter2008}  of NW sector of NGC 6946 showing the HI Northern outer arm. Right: enlarged view on the  the Red Ellipse region according to the KPNO  (top) and Subaru (bottom) telescopes images (optical broad-band data with \Ha-emission in the red channel, see Sec.~\ref{sec_ima}). Orange rectangles   show the regions zoomed in a forthcoming picture. {\it \copyright Subaru telescope and \copyright National Optical Astronomy Observatory/Association of Universities for Research in Astronomy/National Science Foundation}
}
\label{fig_image}
\end{figure*}

The ellipse is assumed to be within NGC 6946, because it is within the chain of HI  clouds, H II regions, and luminous stars  comprising the isolated Northern arm  of this galaxy (Fig.~\ref{fig_image}).  At a distance to NGC 6946 of 5.9 Mpc \blue{\citep[according the updated nearby galaxy catalog by][]{Karachentsev2013}}, the ellipse has a very large size of approximately $170\times290$ pc. \blue{The accepted  distance   is close to the mean value  of a  few dozen measurements of the NGC 6946 distance in the literature, while the most recent values are  $5.86\pm0.76$ Mpc \citep{BoseKumar2014} and $6.72\pm0.15$ Mpc \citep{Tikhonov2014}. }
 In this paper, we present the results of spectroscopic study  of  this elliptical nebulae on the 6-m telescope of the Special Astrophysical Observatory of the Russian Academy of Sciences (SAO RAS).

\section{OBSERVATIONS AND DATA REDUCTION}

\subsection{Archival direct imaging}
\label{sec_ima}
The Red Ellipse was found at the color image presented in  the Subaru 8.2-m telescope press-release\footnote{\url{http://www.subarutelescope.org/Pressrelease/2009/09/08/index.html}}. The data  were obtained by the Subaru Observation Experience Program Team in the period of Aug 28 - Sep 3, 2008 with  the Suprime-Cam imaging camera  and have not been published in  peer-review journals yet. The image  presents an excellent  combination of a large  field ($32\times25$ arcmin) and high spatial resolution (pixel size $0.2$ arcsec). Figure~\ref{fig_image} (bottom right)  shows the fragment  of the Subaru press-release  composite color image:  blue channel is image in the broad $B$-band filter (centered on 0.45 $\mu m$), the green channel is the $V$ filter (centered on 0.55 $\mu m$) image, and the red channel presents data in the IA651 narrow-band  filter centered on 0.651 $\mu m$ (\Ha+[NII]+continuum).  The seeing value estimated as FWHM of single stars in the $V$ band was $0.7$ arcsec.

Later on,  we found that  the enigmatic ellipse is also well seen in  the  color image obtained  at   the  Kitt Peak National Observatory (KPNO) 4-m Mayall telescope  with the Mosaic Camera   on Sep 5, 2008. \footnote{The NOAO image gallery\url{https://www.noao.edu/image_gallery/html/im1092.html}} (Fig.~\ref{fig_image},  top right).   The composite color image presents the $B$ band (blue channel), $V$ band (green channel), and  \Ha+[NII] continuum (red channel) images. The pixel size was similar with the Subaru data ($0.26$ arcsec), while the seeing was $1.0$ arcsec in the $V$ band.

We use the Subaru and KPNO press-release images to consider the morphological structure of the nebulae without photometric estimations. The astrometry grid was created using the Astrometry.net project web-interface \footnote{\url{http://nova.astrometry.net/}}.

The NE part of the Red Ellipse has a short parallel  arc, as well as many other red and white details (Fig.~\ref{fig_image}). The ellipse slightly resembles  a  large supernova remnant somewhat similar to the ultraluminous supernova remnant complex (MF 16) in the NE arm of NGC 6946, studied by \citet*{Blair2001}  with HST WFPC2 images. That ellipse, however, is only about $20\times30$ pc (in spite of being the largest SNR in NGC 6946   known until now)  and has strong X-ray emission associated with it.

In any case, the elliptical ring is not a \blue{young} supernova remnant, because there is no [Fe II] or X-ray emission, as follows from the data shown by \citet{Bruursema2014}. The Red Ellipse is ten times larger and much brighter than the  MF 16, and this  fact alone  casts strong doubt on the identification of the Red Ellipse  as a SNR.

\begin{figure}
\includegraphics[width=0.5\textwidth]{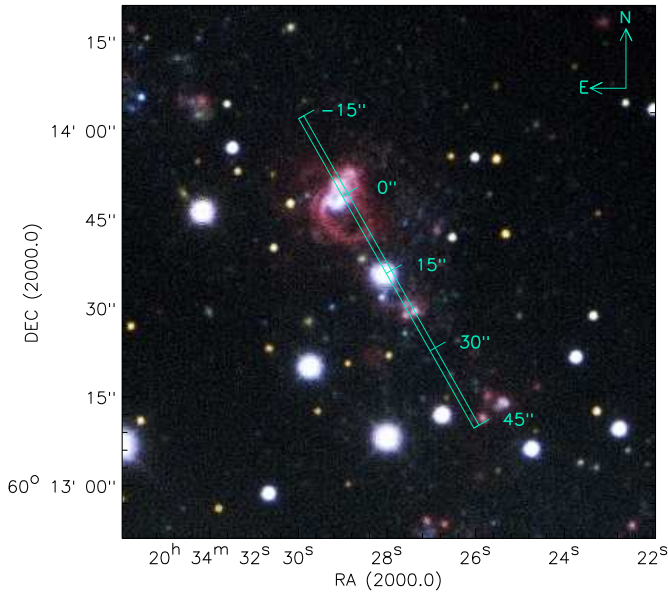}
\caption{The region    of the Northern arm immediately around the Red ellipse (KPNO image).   The position of the SCORPIO-2 slit is overlapped together with the radial scale,  the value $r=0''$ was accepted  to the continuum brightness peak within the slit, i.e., the bright star cluster centre.  }
\label{fig_slitpos}
\end{figure}

\begin{figure*}
\includegraphics[width=0.95\textwidth]{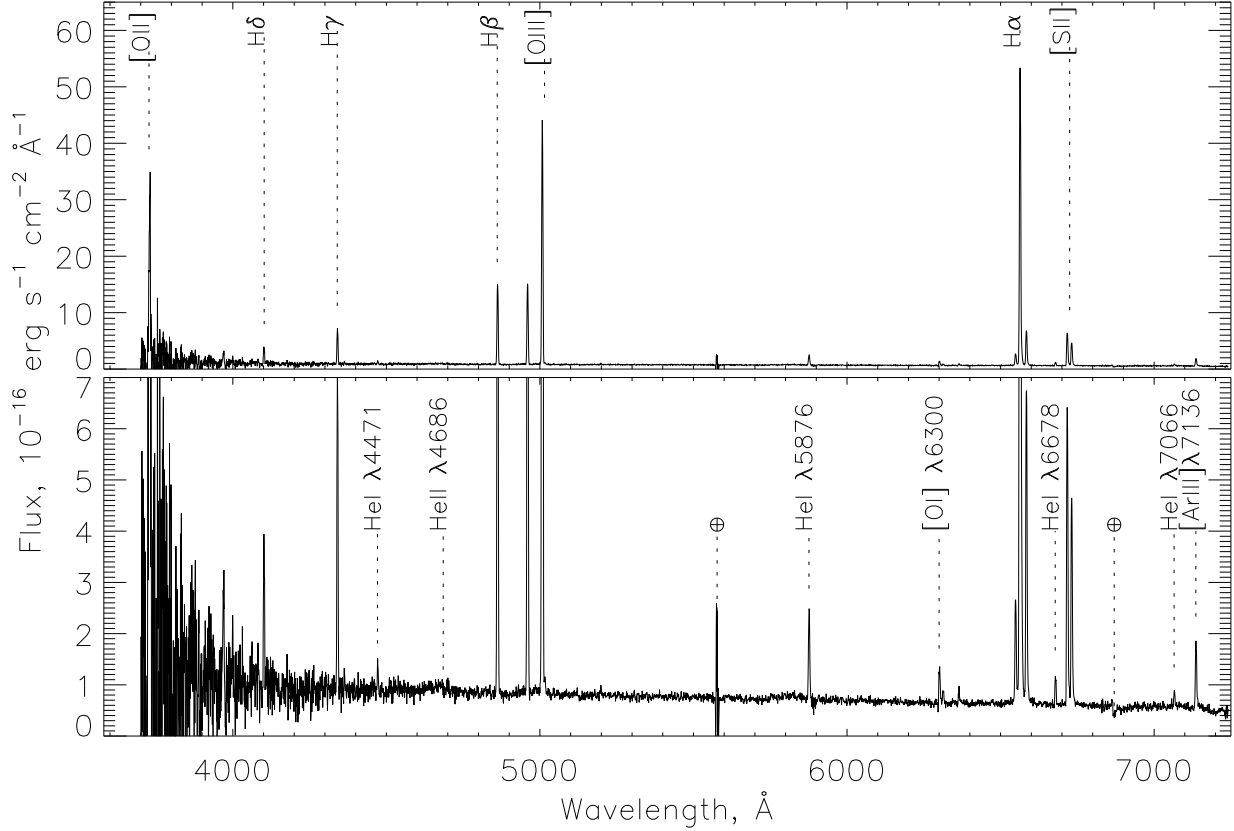}
\caption{Integrated spectrum of the  brightest part of the nebula $\pm3$ arcsec around the star cluster centre in two scales of intensity. The brightest  and weak emission lines (together with remnants of air-glow emissions) are labeled on top and bottom plots correspondingly.}
\label{fig_spectrum}
\end{figure*}

\subsection{Spectroscopy}

Long-slit spectral observations were  carried out at the prime focus of the SAO RAS  6-m Big Telescope Alt-azimuthal (BTA) with the SCORPIO-2  multimode focal reducer \citep{AfanasievMoiseev2011} in the night  of 6/7 November 2015. The slit was 6.1 arcmin in length and 1 arcsec in width, the  grism VPHG1200@540 provided a spectral resolution  of about 4.5-5\AA\,  in the  spectral range of 3650--7250\AA\, with  a mean reciprocal dispersion of 0.8\AA\, per pixel, and with a                spatial sampling of 0.36 arcsec per pixel along the slit. The total integration time was 2400 sec under  seeing $1.7-1.8''$.  We  put  the slit across the nebula major axis ($PA=210\degr$),  see Fig. \ref{fig_slitpos}. The object data and spectrum of the spectrophotometric standard star observed at the same night  were reduced and calibrated using the IDL-based software developed at the SAO RAS, \citep*[see, for instance,][]{Egorov2013}

The parameters of the emission lines (integrated  flux, line-of-sight velocity, FWHM) were calculated from  a single-gaussian fitting  using a similar technique  described in \citep{Egorov2013}. The doublets ([NII]$\lambda6548,6583$, [OIII]$\lambda4959,5007$, and [SII]$\lambda6717,6731$) were  fitted by a pair of Gaussians  with the same velocities and FWHM. We estimated the errors of the measured parameters  by analyzing the synthetic spectra degraded by noise with the same  signal-to-noise ratio as in the corresponding observational spectrum. To increase the signal-to-noise ratio for weak emission regions, we binned spectrograms into 2 pixels bins along the slit prior to emission line analysis. The final spatial scale  was 0.7 arcsec per pixel in a good accordance with the atmospheric seeing value.
A multicomponent structure of the emission-line profiles was not detected. Also,  the FWHM of the main emission lines is  in a good agreement with the width of the nearest airglow emission lines, we can only conclude that the actual (free from instrumental contour broadening) FWHM of the lines is smaller than 50--70$\km$.

\section{\blue {RESULTS}}

\subsection{\blue{The nebulae integrated spectrum.}}
\label{sec_intspec}

Figure~\ref{fig_spectrum} shows the integrated spectrum around ($\pm3''$) of the star cluster. The spectrum appears to be typical of an HII region; however, weak helium lines are also detected  including the low-contrast broad bump around HeII$\lambda4686$.  Therefore, some emission-line hot stars like Of or even WR could belong  to the cluster. Noise  on the spectrum shown on Fig.~\ref{fig_spectrum}  grows up significantly  at the short wavelength, \blue{because the quantum efficiency of CCD E2V42-90 detector decreases  sharply in the blue part of the spectrum.}.

\blue{
Table~\ref{tab_ratio} lists  the  observed  intensity of emission lines ($I(obs)$) in  the integrated spectrum shown in Fig.~\ref{fig_spectrum} derived from the Gaussian fitting, as was described in previous section. The intensities are given in a standard scale where $I(\mbox{H}\beta)=100$, here and bellow the errors correspond to $3\sigma$  level. We determined the $E(B-V)$ color excess  from the  Balmer decrement using the theoretical intensity ratios of the hydrogen lines from the electron temperature $T_e=10\,000$~K \citep[Case B recombination, see][]{Osterbrock1989} and the \citet{Fitzpatrick1999} parametrization of the \citet*{Cardelli1989} extinction  law.   We obtained the following values for the color excess:
}

\blue{
$E(B-V)=0.302\pm0.006 $  from \Ha/$\mbox{H}\beta$ flux ratio,
}

\blue{
$E(B-V)=0.356\pm0.075$  from $\mbox{H}\gamma/\mbox{H}\beta$ ratio,  and
}

\blue{
$E(B-V)=0.694\pm0.212 $  from $\mbox{H}\delta/\mbox{H}\beta$ ratio.
}

\blue{
For dereddening of observed line intensity we used the first value $E(B-V)=0.302$ because it has the significantly smaller  uncertainty. The  reddening-corrected intensities    $I(red. corr.)$ are also  shown in Tab.~\ref{tab_ratio}. The accepted $E(B-V)$ value corresponds to the visual extinction $A_V=0.94\pm0.02$ ($R_V=3.1$) in a nice agreement with the value $A_V=0.938$  presented in  NASA/IPAC Extragalactic Database for coordinates of NGC6946  according  Galactic extinction data from  \citet{SchlaflyFinkbeiner2011}. It evidens that the extinction observed in direction of the Red Ellipse is caused by our Galaxy.
}

\blue{
We calculated the nebular  electron temperature $T_e(OIII)$ from the [OIII]($\lambda4959$+$\lambda5007$)/[OIII]$\lambda4363$ lines intensity ratio according the improved calibration relation of \citet{Pilyugin2010}:  $T_e(OIII)=9900\pm1800$~K. The relative high  uncertainty in $T_e$ is produced by  large errors in [OIII]$\lambda4363$ flux estimation.  The average electron density $n_e$ derived from the [SII] $\lambda6731/\lambda6717$ line ratio according  \citet{Osterbrock1989}  for $T_e=10\,000$~K  is $n_e<10\,\mbox{cm}^{-3}$.}

\begin{table}
\blue {    \caption{Relative intensities of emission lines in the brighter part of the Red Ellipse [$I(\mbox{H}\beta)=100$].}
    \label{tab_ratio}
    \begin{tabular}{lrr}
        \hline
        Lines        & $I(obs)$ & $I(red. corr.)$ \\
        \hline
    [OII]$\lambda 3727,3729$ & $258.0\pm50.9$ & $345.4\pm68.2$ \\
 H$\delta$ $\lambda 4104$ & $ 16.7\pm 2.3$ & $ 20.3\pm 2.8$ \\
 H$\gamma$ $\lambda 4340$ & $ 39.9\pm 1.3$ & $ 45.9\pm 1.5$ \\
  \,   [OIII] $\lambda 4363$             & $  1.7\pm 1.0$ &   $  1.9\pm 1.2$ \\
    HeI$\lambda 4471$ & $  2.6\pm 0.8$ & $  2.9\pm 0.9$ \\
   HeII$\lambda 4686$ & $  1.1\pm 1.1$ & $  1.2\pm 1.2$ \\
  H$\beta$ $\lambda 4861$ & $100.0\pm 0.7$ & $100.0\pm 0.7$ \\
\,[OIII]$\lambda 4959$ & $101.2\pm 0.2$ & $ 98.5\pm 0.2$ \\
\,[OIII]$\lambda 5007$ & $308.0\pm 0.7$ & $295.9\pm 0.6$ \\
    HeI$\lambda 5876$ & $ 13.3\pm 0.5$ & $ 10.5\pm 0.4$ \\
\,   [OI]$\lambda 6300$ & $  5.4\pm 0.5$ & $  3.9\pm 0.4$ \\
\, [NII]$\lambda 6548$ & $ 16.3\pm 0.2$ & $ 11.6\pm 0.1$ \\
  \Ha~$\lambda 6563$ & $405.6\pm 0.7$ & $286.0\pm 0.5$ \\
\, [NII]$\lambda 6583$ & $ 48.0\pm 0.6$ & $ 33.7\pm 0.4$ \\
    HeI$\lambda 6678$ & $  3.8\pm 0.3$ & $  2.7\pm 0.2$ \\
\,  [SII]$\lambda 6717$ & $ 45.2\pm 0.4$ & $ 31.2\pm 0.3$ \\
\, [SII]$\lambda 6731$ & $ 31.5\pm 0.3$ & $ 21.6\pm 0.2$ \\
    HeI$\lambda 7065$ & $  1.9\pm 0.3$ & $  1.2\pm 0.2$ \\
\,[ArIII]$\lambda 7136$ & $ 10.5\pm 0.4$ & $  6.8\pm 0.3$ \\
            \hline
    \end{tabular}
}
\end{table}

\begin{figure}
\includegraphics[width=0.5\textwidth]{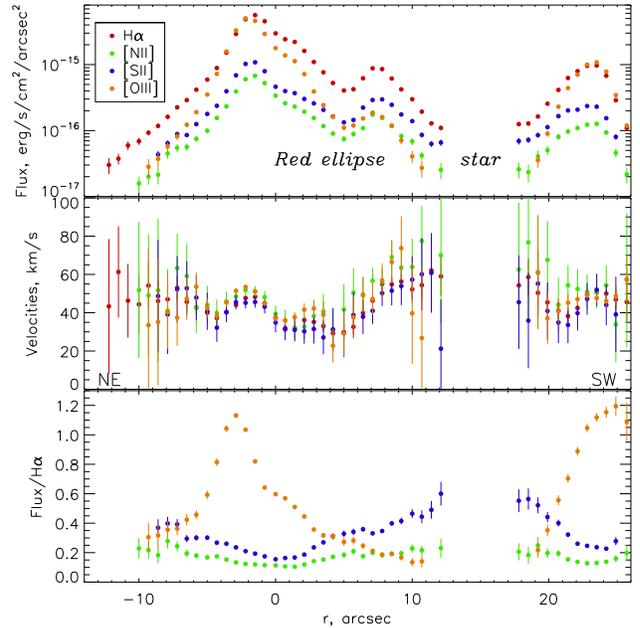}
\caption{ Parameters of the brightest emission lines along the slit (Gaussian fitting results): the surface  brightness (top),  line-of-sight velocities (middle), and flux ratio relative to \Ha-intencity (bottom). The error-bars correspond to  $3\sigma$ level.} \label{fig_lines}
\end{figure}

\begin{figure*}
\centerline{\includegraphics[width=\textwidth]{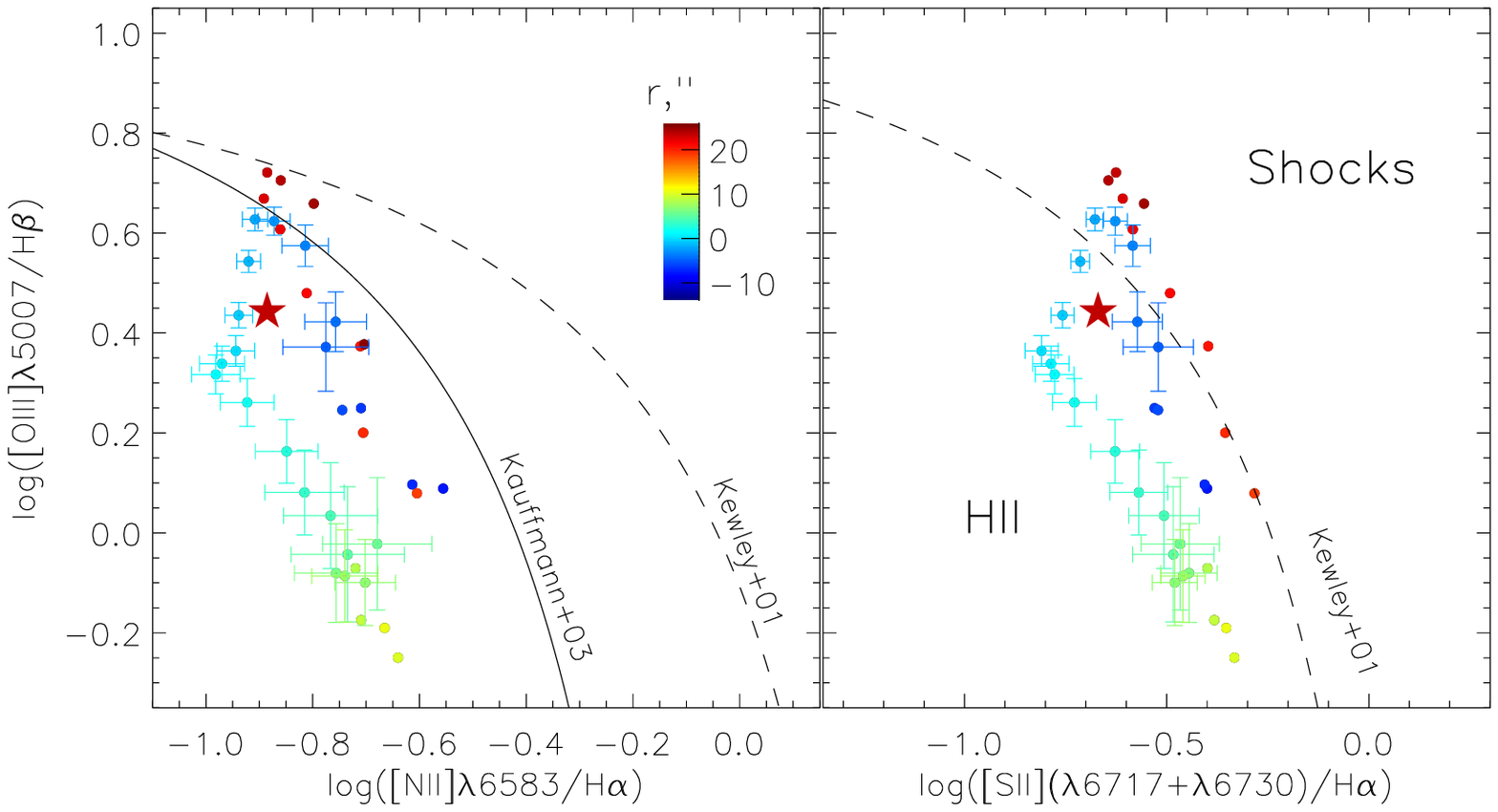}}
\centerline{\includegraphics[width=\textwidth]{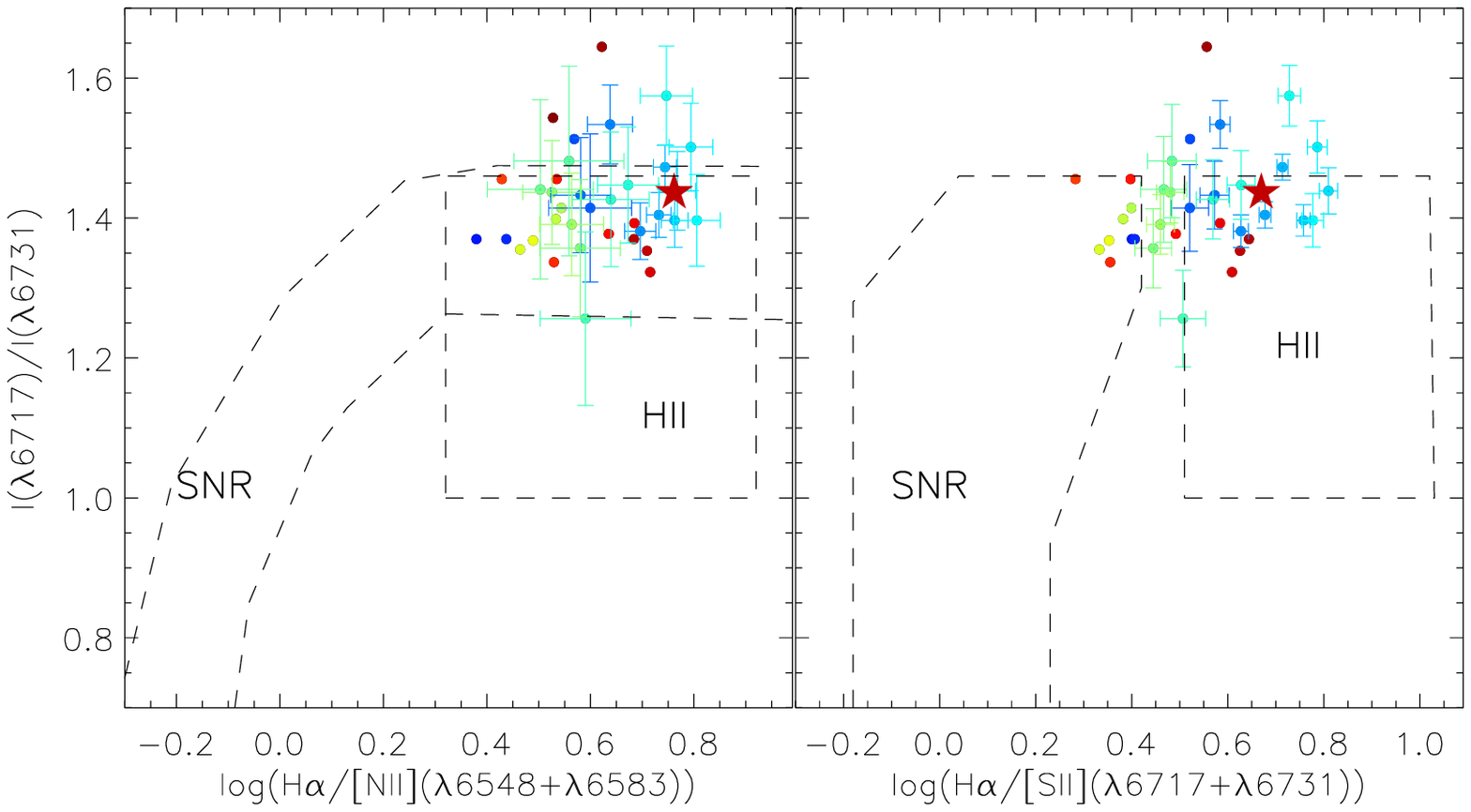}}
\caption{Excitation diagnostic diagrams, comparing the emission-line intensity ratios. The points color corresponds to different locations along the slit according to the scale-box. The $3\sigma$ error-bars are shown for the points in the main part of the Red Ellipse ($r=-5...+8''$). The top panels present  classical BPT diagrams.  The dividing lines between the regions ionized by hot stars and shocks/AGN are shown according to \citet{Kewley2001} and \citet{Kauffmann2003}.
The bottom panels show diagrams used for separation between SNR and HII regions in nearby galaxies with  the dividing borders adopted from  \citet{Leonidaki2013}.
\blue{The red star marks the lines ratio derived from the total spectrum of the nebulae integrated in the range $r=-5...+8$ arcsec.}
}\label{fig_ratio}
\end{figure*}

\begin{figure*}
\includegraphics[width=\linewidth]{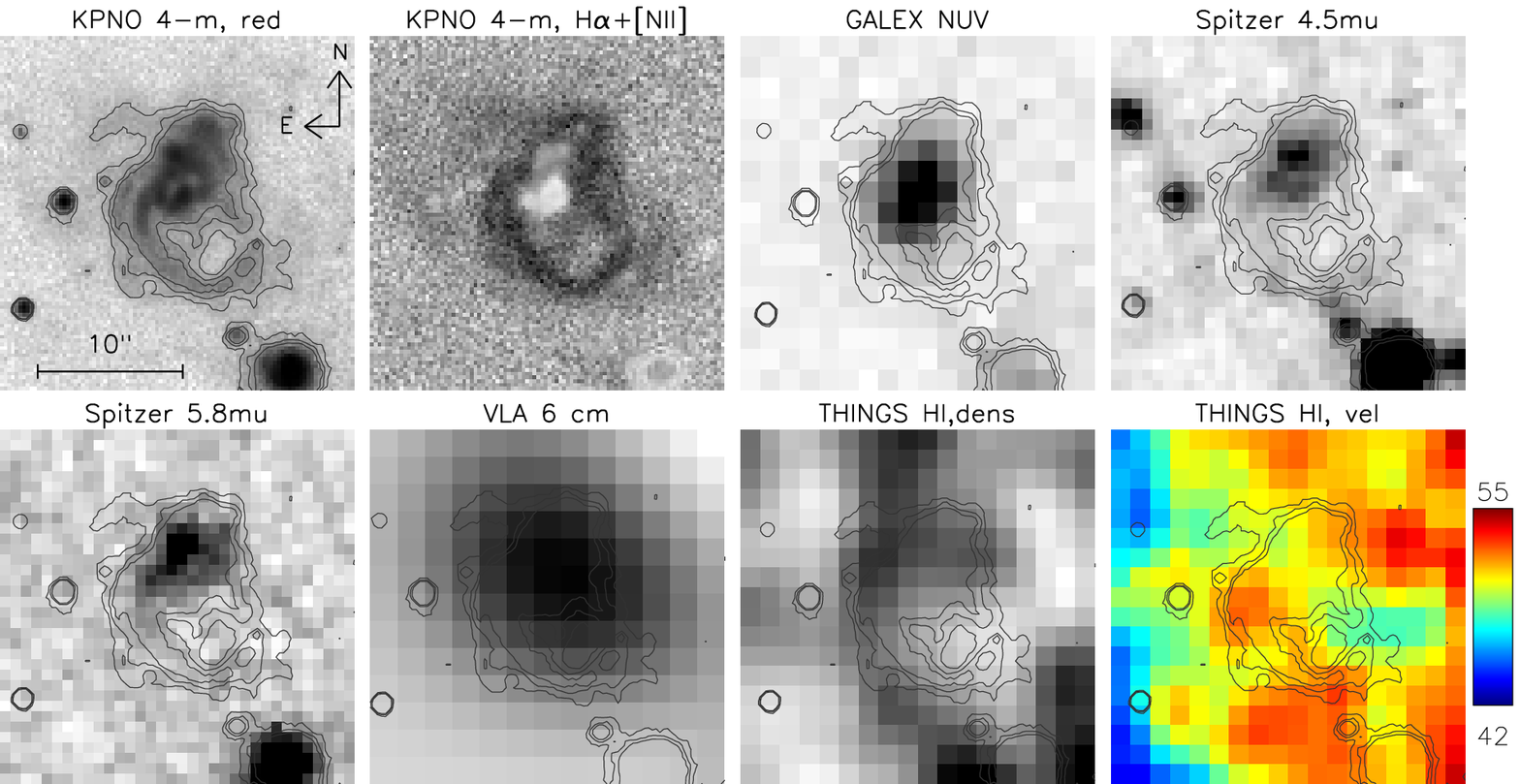}
\caption{Collection of multi-wavelength data for  the Red Ellipse: KPNO image (emission lines and continuum), the net \Ha+[NII] image, \textit{GALEX} near-ultraviolet image, \textit{Spitzer} SINGS near-infrared  data at 4.5 and 5.8 $\mu$m, VLA 5GHz radio continuum \citep{Beck1991},  THINGS \citep{Walter2008} HI column density with $6''$ resolution and velocity field (the first moment map) scaled in $\km$. Three isophotes of the KPNO red image are overlapped in order to show the nebula location.}\label{fig_multiwave}
\end{figure*}

\subsection{\blue {Emission-lines kinematics }}

Only a few of the brightest emission lines are detected in the nebula outside the star cluster regions including the Balmer (\Ha, H$\beta$) and forbidden ([SII]$\lambda6717,6731$, [NII]$\lambda6548,6583$, [OIII]$\lambda4959,5007$) lines. Changes of their parameters across the slit are shown in Fig.~\ref{fig_lines}. The coordinates along the slit ($r$)  are positive in WS directions with the coordinate origin in the star cluster photometric centre according to the ruler denoted in Fig.~\ref{fig_slitpos}. Hereafter, we use a term ``the main part'' of the nebulae/Red Ellipse for  points in the coordinate range $r=-5...+8''$ corresponding to the bright borders of the objects in \Ha-images, see Figs.~\ref{fig_image} and \ref{fig_slitpos}. Also, our  spectrograms allow us to study weak extended diffuse gas around the Red Ellipse. Note a significant  light pollution from the Galaxy foreground star at $r=13-17''$ near the SW border of the Red Ellipse. We can make the following conclusions based on the changes of emission-line parameters along the slit:

\begin{itemize}

\item[(i)] The velocities of the forbidden and Balmer lines are in a good agreement; their values across the  Red Ellipse nebula lie in the range of 30--60$\km$ that is in a good agreement with the HI observations (45--55$\km$ , see below Sec.~\ref{sec_discussion}).  If velocity variations in the main part of the Red Ellipse ($r=-2...+8''$) are caused by nebula expansion, the corresponding expansion velocity should be less than 15--20$\km$.

\item[(ii)]Behaviors of the forbidden-to-recombination lines ratio are different in the different lines. The relative flux in   [OIII]$\lambda5007$ increases in the NE side of the Red Ellipse (around the star cluster), while it  decreases to the SW arcs,  where the  [SII]/\Ha\, and [NII]/\Ha\, ratios grow (hereafter, we use [N II]/\Ha\,and [S II]/\Ha\,  to denote the [N II]$\lambda6583$/\Ha\, and [S II]($\lambda6717+\lambda6731$)/\Ha\, line flux ratios respectively). Obviously, the growth of a  temperature-sensitive [O III] emission line is related to the ionization radiation of hot massive stars in the cluster.
\end{itemize}

\subsection{\blue{Gas ionization properties}}

We used  the ratios of neighboring emission lines  independent  of interstellar extinction    to probe  the gas ionization state. All points in the main part of the nebulae satisfy the case of photoionization  by young stars according to the popular criteria  $ [S II]/$\Ha$<0.4$ and $[N II]/$\Ha$ < 0.4$ \citep{Stasinska2006}. However these ratios exceed 0.4--0.5 in diffuse gas outside the bright nebulae ($r>10''$, see Fig.~\ref{fig_lines}). Similar behaviors of gas excitation inside/outside the Red Ellipse are also seen in the classical  BPT diagram \citep*{BPT} present $[S II]/$\Ha\, and $[N II]/$\Ha\, versus $ [OIII]/\mbox{H}\beta$ ratio (Fig.~\ref{fig_ratio}, top panels).
Indeed, most points which belong to the Red Ellipse are located below the conventional lines  \citep{Kewley2001,Kauffmann2003} separated gas photoionized   in the HII-regions and the  regions, where gas is ionized by shocks or by hard UV photos from active galactic nuclei (AGN).  While the points corresponding to the diffuse gas surrounding the main nebulae are located on the border dividing the different  excitation mechanisms or even above the line, i.e., in the shock/AGN region. This fact indicates  that  shocks   contribute to the gas ionization, which is not surprising for  warm diffuse gas surrounding sites of current star formation.  Also, the border line is occupied by points  with $r<0''$ from the vicinity of the star cluster, where the densest cloud of neutral gas is located according to the HI column density map (see below Sec.~\ref{sec_discussion} and  Fig.~\ref{fig_multiwave}). As it was mentioned above, this region is characterized by relatively bright [O III] emissions, indicating high temperature of gas ionized by hot massive stars including shocks generated by stellar winds from WR and Of emission-line stars.

Also, we considered diagnostic diagrams separating different types of nebulas (SNR, HII-regions, and PN) using  electron density-depended ratio of sulfur dublet components [SII] $I(\lambda6731)/l(\lambda6717)$  versus \Ha/[NII] and \Ha/[SII] lines ratio  \citep*{Sabbadin1997,Leonidaki2013}. Bottom panels in Fig.~\ref{fig_ratio} show that the points from the main part of the Red Ellipse avoid  the region of Galactic and extragalactic SNR. These points are  within  the range of the lines ratios  which is corresponded to  HII-regions and  the border SNR/HII. The observed data are distributed along vertical axis in the directions of   high ratio  $I(\lambda6731)/l(\lambda6717)=1.30-1.44$ corresponding  to the electron  densities $n_e=0-120\,\mbox{cm}^{-3}$ \citep[for $T_e=10\,000$~K according][]{Osterbrock1989}. Some points are located  even above the line ratio 1.44 that is formally forbidden by this `standard'  $I(\lambda6731)/l(\lambda6717)$--$n_e$ relation. It implies that  the mean $n_e$ in the Red Ellipse is smaller then in the typical HII regions, \blue{in agreement with our estimation for the integrated spectrum of the Red Ellipse ($n_e<10\,\mbox{cm}^{-3}$, see  Sec.~\ref{sec_intspec}). }This fact together with the ionization  by hot stars  and small contribution of shock points to superbubble  created by collective actions of massive stars in the NE star cluster, as the most likely explanation of the Red Ellipse origin.

\blue{As it follows from Figs.\ref{fig_lines} and \ref{fig_ratio}, the [SII]/\Ha\, lines ratio exceeds the  conventional value  0.4 (which separated shocks and photoionization)  in the faint arc at the SW border of the nebula ($r=+5...+8''$).  Therefore, the shock-heated  gas make  some contribution to the observed optical spectrum. The shocks might be generated by explosions of SN related with the central star cluster. However, their possible contribution  to the total ionization  seems to be negligible, because the lines ratios for the integrated nebular spectrum  corresponds to massive stars photoionization (see the red stars  in Fig.~\ref{fig_ratio}).  Also, the value of electron temperature estimated for the Red Ellipse ($T_e(OIII)=9900\pm1800$~K, sec.~\ref{sec_intspec}) is significantly low comparing with  $T_e$ observed in SNR in local galaxies \citep*[see, for example, ][]{Galarza1999}. Therefore the Red Ellipse can not be considered as a `giant SNR' created by a single SN explosion.
}

\section{DISCUSSION}
\label{sec_discussion}

\citet{Ferguson1998} were among the first to study star formation in the extreme outer regions of disc galaxies including NGC 6946. Their  \Ha-image shows the  region of the Red Ellipse as an unresolved bright \Ha\, source. They infer that the star formation rate is fairly constant from the inner to the outer disc in NGC 6946, although, the covering factor of star formation in the outer disc is quite low, so the overall azimuthally averaged star formation rate abruptly drops in the outer regions.

Nevertheless, the outer northern spiral arm containing the Red Ellipse has the highest HI column density of any region in the outer galaxy,  with a value of $1.5\cdot10^{21}\cm$ \citep{HaanBraun2014}. It also has an unusual inward bulk motion of 30$\km$, which led \citet{HaanBraun2014} to suggest that this arm was compressed and moved by ram pressure from an intergalactic cloud. This cloud appears in the HI map of \citet{Pisano2014}.

We might infer that this ram pressure has triggered the super star cluster, which resulted in supernovae leading to formation of a giant (even supergiant for a SNR) elliptical ring; anyway, we know that SNe may arise in a poor cluster as well (such as SN 1987a, the probable member of a small open cluster in the LMC \citep{Efremov1991}. The most intriguing feature of the nebula is its  very regular  elliptical shape similar to some SNR or Galactic  ring nebulae around WR or Of stars. However, the size of the Red Ellipse ($\sim300$ pc along major axis) is comparable to wind blown structures related to giant HII-regions. Other arguments against SNR origin of the Red Ellipse are the following: (i) the lack of  the X-ray and synchrotron radio emission; (ii) the contribution of shocks in  gas ionization is very modest,  as follows from our diagnostic diagrams. The multi-wavelength archive  images of the nebula (Fig.~\ref{fig_multiwave}) clearly demonstrate that \textit{GALEX} UV, \textit{Spitzer} MIR, and 6-cm radio emission are directly related to the star cluster in the base of the Red Ellipse. The HI line-of-sight velocities are  in a good agreement with our estimation in the ionized gas if the difference of spatial resolutions will be taken into account. The HI velocity variations through the nebulae are modest and do not exceed a few $\km$. The extended diffuse regions of the nebulae are not detected in the present multi-wavelength data set with  exception of  \textit{Spitzer} data. Indeed, \textit{Spitzer} 4.5 and 5.8 $\mu m$  images reveal  weak  counterparts in SW side of the Red Ellipse, where the cavern of low-density gas appears on the HI density map. We suggest that this feature could be related with  warp dust emission,  including Polycyclic Aromatic Hydrocarbons contributions which are usually observed in HII regions of active star formation.

\section{CONCLUSIONS}

The Red Ellipse seems to be a nebula created by  collective actions of SN explosions  and  winds of massive stars in the NE stellar cluster---the presence of WR or Of stars is evident on the star cluster  spectrum. However, even in this case, the regular and closed morphology of the Red Ellipse is unusual. It is possible that  we observe a rare interplay between pre-existed distribution of cold and dense interstellar medium  and hot gas ejected from young stellar group. Indeed, the cavern in the HI distribution according to the THING data matches the emission-line bubble (Fig.~\ref{fig_multiwave}).  A ``wall'' with the HI surface density  observed near the SW border of the Red Ellipse can prevent its further expansion. A similar (but not the same) example is the well-known S3 giant bipolar nebula (two bubbles about 100 and 200 pc in diameter) created by a single WO star in the dwarf galaxy IC 1613.  \citet{Lozinskaya2001} have shown that the unique structure of S3 is caused by stellar wind of the WO star embedded in a dense layer at the boundary of HI  supercavity.   Does the similar  scenario  work  also in the case of the Red Ellipse? New observations, first of all integral-field data in the optical or/and NIR ranges together with the detailed simulations of feedback processes on the outskirts  of NGC 6946 are crucial for the final conclusion.  The strange \blue{regular} shape of this  nebula remains enigmatic.

\section*{ACKNOWLEDGMENTS}
The observations at the 6-meter BTA telescope were carried out with the financial support of the Ministry of Education and Science of the Russian Federation (agreement No. 14.619.21.0004, project ID RFMEFI61914X0004). In this  research, we  used  the NASA/IPAC Extragalactic Database (NED) which is operated by the Jet Propulsion Laboratory, California Institute of Technology under contract with the National Aeronautics and Space Administration.  The \textit{GALEX} image presented in this paper was obtained from the Mikulski Archive for Space Telescopes (MAST). STScI is operated by the Association of Universities for Research in Astronomy, Inc., under NASA contract NAS5-26555. The support for MAST for non-HST data is provided by the NASA Office of Space Science via grant NNX09AF08G and by other grants and contracts. We also note  very  useful  collection of the NGC 6946 images obtained at  the KPNO  on the Mayall  4-m telescope with  the Mosaic camera.  This study has been  carried out under the financial support of  the Russian Science Foundation (project No. 14-22-00041).  We are very grateful to Debra Elmegreen and Bruce Elmegreen  for  useful  discussions and also  to the anonymous referee for the constructive comments and suggestions that helped us to improve and clarify our result.  We thank Roman Uklein and Dmitri Oparin for assisting in the 6-m telescope observations.

{}

\label{lastpage}

\end{document}